\documentclass[
    ,final            
  ,numberedheadings 
  ]
  {aipproc}

\layoutstyle{6x9}

\begin{document}

\title{Criterium for the index theorem on the lattice}

\author{Pedro Bicudo}{
  address={Dep. F\'{\i}sica and CFIF, Inst. Sup. T\'ec., 1049-001 Lisboa, Portugal}
}

\begin{abstract}
We study how far the Index Theorem can be extrapolated
from the continuum to finite lattices with finite topological charge
densities. To examine how the Wilson action approximates the
Index theorem, we specialize in the lattice version of the
Schwinger model. We propose a new criterion for solutions of 
the Ginsparg-Wilson Relation constructed with the Wilson action. 
We conclude that the Neuberger action is the simplest one that 
maximally complies with the Index Theorem, and that its best 
parameter in $d=2$ is $m_0=1.1 \pm 0.1$ .
\end{abstract}

\maketitle


\section{introduction}
\label{sec-introduction}

\par
One of the constraints that lead to the Nielsen-Ninomiya
\cite{Nielsen} no-go theorem, is the chiral invariance of the Dirac 
action $D$ for massless fermions on the lattice. Under certain 
conditions \cite{Nielsen} , the spectrum of fermions $\psi$ would 
suffer from doubling, and the axial 
anomaly would cancel. This has been, from the onset, a recurrent
problem of Lattice QCD \cite{Wilson}.
In order to recover the axial anomaly, at least at the pertubative 
level, Ginsparg and Wilson \cite{Ginsparg} derived
a relation, the Ginsparg-Wilson Relation (GWR) which 
explicitly breaks the standard chiral symmetry,
\begin{equation}
\label{GWR 5}
D \, \gamma_5 + \gamma_5 \, D = D \, \gamma_5 \, R \, D \ ,
\end{equation}
where $R$ is a matrix proportional to the lattice spacing $a$
(in most of this talk we will consider the case where
$R$ is a simple number).
Recently L\"usher \cite{Luscher} proved that chiral
invariance of an action $D$ which complies with the GWR
can be recovered in an extended form,
\begin{equation}
\delta \psi = \gamma_5 (I -{1 \over 2} \, R D ) \psi \ , \ \
\delta \bar \psi = \bar \psi (I -{1 \over 2} D R ) \gamma_5 \  .
\label{chiral rotation}
\end{equation}
Thus a conjecture appeared in the literature suggesting that it 
may be possible to overcome the Nielsen-Ninomiya no-go theorem on
the lattice and to fully simulate, without doubling, the chiral
symmetry of QCD (see \cite{Niedermayer} for a recent review).
This 
includes PCAC \cite{Chandrasekharan} and the axial anomaly
\cite{anomaly}. 
In particular the Atiyah-Singer \cite{Atiyah} Index Theorem  
has a Hasenfratz-Laliena-Niedermayer \cite{Hasenfratz-index}
version on the lattice,
\begin{equation}
\label{index}
n_- -n_+ = q \ ,
\end{equation}
where $n_{-(+)}$ is the number of zero modes of the Dirac action 
$D$ with negative (positive) chirality. $q$ is 
an {\em integer} 
topological charge of the gauge field configuration, defined with
the charge density, a function of the lattice site $\vec n$,
\begin{equation}
\rho(\vec n)={1 \over 2}\, tr
{ \atop \Biggl|}
\begin{array}[t]{l}
\ \ \{ \gamma_5 \, R \, D(\vec n, \vec n)\} 
\\ _{colour, \, Dirac}
\end{array}
\ .
\label{GWR density}
\end{equation}
Therefore each GWR Dirac action {\em defines} the 
density (\ref{GWR density}) and the index (\ref{index}).
Moreover examples of lattice actions have been found
where this index is nontrivial and coincides with
a charge independently defined from a topological 
gauge density. 
For instance the fixed point topological density 
\cite{Hasenfratz-fixed} 
coincides, on all gauge configurations, with the density 
(\ref{GWR density}) 
defined with the fixed point Dirac operator. 
The recent overlap GWR solution of Neuberger \cite{Neuberger} 
also solves the GWR and complies with the Index Theorem.
The Neuberger solution was checked in dimension 
$d=2$ for smooth gauge configurations 
\cite{Chiu-testexact,Chandrasekharan-Schwinger} ,
and in dimension $d=4$ in the continuum limit \cite{Fujikawa,Adams}. 
Thus the GWR, together with these recent results, essentially
solve the problem of the topological Index Theorem on the lattice. 
\par
Nevertheless we remark that the definition of the Dirac operator 
and the definition of the topological charge on the lattice are not unique.
Moreover the topological charge on the lattice is not conserved \cite{'tHooft}.
This suggests that different GWR Dirac operators and topological densities  
produce different indexes when we depart from the continuum limit, to a 
finite lattice with finite charge density. In this talk we address the 
{\em extremum} problem of finding the lattice Dirac operator which provides the 
best index (\ref{index}). 
Because this is a difficult problem, we specialize in 
lattice actions which are constructed with the Wilson action 
\cite{Wilson,Gattringer}.
In order to compare the index of different Dirac 
operators, we first choose a single topological charge density.  
Although this is an arbitrary definition, let us choose the simplest near 
neighbor lattice topological charge density which discretizes the continuum
density $\epsilon_{\mu\,\nu \cdots}F^{\mu\, \nu}\cdots$ and which sums 
to an integer total topological charge. 
Then we search for the GWR Dirac operator which 
index (\ref{index}) is closer to this simplest topological charge.  
\par
The arbitrariness in the choice of our lattice operators is partly 
constrained by locality. 
Locality is a crucial issue to connect the 
lattice to the Quantum Field Theories in the continuum. 
Only local actions are universal. 
Moreover it has been conjectured  \cite{Niedermayer,local} that local 
GWR solutions have better topological properties than non-local 
ones. Only topological configurations which are larger than the range of the 
couplings in $D$ would produce a number of zero modes $n_-$ different from 
$n_+$ in agreement with eq. (\ref{index}). This conjecture is certainly correct
in the case of the Zenkin \cite{Zenkin,Chiu-method} action. The Zenkin action
is non-local \cite{Bicudo}, and although it complies with eq. (\ref{GWR 5}),
it does not comply with eq. (\ref{index}). 
Thus we will restrain from using non-local definitions for the lattice 
action and for the lattice charge density.
\par
In this talk we investigate in detail the eigenvalues
of some lattice actions, with the aim to clarify how 
the Index Theorem is extrapolated from the continuum 
to finite lattices with finite topological charge
densities. We also propose a new criterion for 
GWR solutions constructed with the Wilson action. 
In Section \ref{sec-eigen Wilson} we study in detail the effects of
topological gauge configurations on the eigenvalues
of the Wilson action for the discrete version 
of the Schwinger model. 
In Section \ref{sec-properties} we review properties of lattice actions.
In Section \ref{sec-topology} we address the topological Abelian charge in $d=2$.
In Section \ref{sec-Wilson} we review the Wilson action.
In Section \ref{sec-criterion} we present our criterion.
Finally in Section \ref{sec-conclusion} we conclude.

\section{Some properties of lattice actions}
\label{sec-properties}
\par
A plausible property of lattice actions is $\gamma_5$ hermiticity. 
For instance the Wilson action is $\gamma_5$ Hermitean.
When an action verifies $ D^\dagger=\gamma_5 \hspace{-.05cm}D \gamma_5$,
we can show that $D$ and $D^\dagger$ have the same eigenvalues $\lambda$,
\begin{eqnarray}
D \ v &=& \lambda \  v \nonumber \\
\Leftrightarrow  \ \
D^\dagger \ \gamma_5 \ v &=& 
\lambda \  \gamma_5 \ v   \ ,
\label{sameig}
\end{eqnarray}
and that the eigenvalues of $D$ and $D^\dagger$ are mapped by $\gamma_5$.
When we conjugate eq (\ref{sameig}), we find that for every right 
eigenvalue there is a left eigenvalue $\lambda^*$ with eigenvector
$v^\dagger \ \gamma_5$, 
and so the conjugate transformation 
$\lambda \rightarrow \lambda ^*$ leaves the spectrum invariant.
The spectrum of eigenvalues $\lambda$ is symmetric with respect to 
the real axis of the Argand plot.
\par
We proceed to define chirality in the lattice.
Eq. (\ref{sameig}) can be used to show that the $\gamma_5$ is involved in 
a extended orthogonality condition,
\begin{equation}
\left(\lambda_2^* - \lambda_1 \right) \, 
v_2^\dagger \ \gamma_5 \ v_1 =0 \ .
\label{orthogonal}
\end{equation}
A possible definition for the chirality of an eigenvector $v$ is,
\begin{equation}
\chi=  { v^\dagger \ \gamma_5 \ v \over v^\dagger \ v} ,
\label{chirality}
\end{equation}
and eq. (\ref{orthogonal})  shows that the complex eigenvalues have a
vanishing chirality, $\chi=0$. Only the real eigenvalues may have
a non vanishing chirality.
\par
We now study properties of GWR solutions.
The GWR implies that if $v$ is a zero mode of the GWR 
action $D \, v=0$, then $D \, \gamma_5 \, v=0$. Thus we 
can use $1-\gamma_5$ and $1+\gamma_5$ to decompose the Kernel in 
a set of left vectors and a set of right vectors. This shows that the 
zero modes have a chirality $\chi=\pm 1$.
Moreover we can verify that if $D$ is a GWR action, then
$-D+ 2/R$ also complies with the GWR. Thus the 
eigenvalues $2/R$ of $D$ also have a chirality $\chi=\pm 1$.
\par
It is also convenient to define an intermediate matrix,
\begin{equation}
\label{intermediate V}
V=R D-I    \Leftrightarrow D = {1 \over R}(I+V)
\end{equation}
because the GWR is equivalent to $V^{-1}=\gamma_5 V \gamma_5$.
In the case that $V$ is $\gamma_5$ Hermitean, we also
obtain that  $V^{-1}= V^\dagger$, and this
shows that the eigenvectors of $V$ belong
to the unitary circle of the complex Argand plane. 
The eigenvalues of $D$ are in a circle with center
$1/R + 0\, i$ and radius $1/R$.
We will be interested in the $-1$ eigenvalues
of $V$ which correspond to zero modes of $D$.
\par
Moreover the crucial property for the Index Theorem has been show,  
\begin{equation}
{R \over 2} tr\{ \gamma_5 D \}= n_- -n_+ \, ,
\end{equation}
both with analytical \cite{Luscher} and algebraic \cite{Fujikawa} methods.

\section{Topological configurations in d=2}
\label{sec-topology}
The $d=2$ case is very convenient for the simple
study of discrete topology. The Euclidean Dirac
matrices $\gamma_{1 \rightarrow 5}$ of $d=4$ are now replaced 
by the Pauli matrices $\sigma_{1 \rightarrow 3}$. 
The $d=2$
charge density is similar to the rotational of 
the vector potential,
\begin{equation}
\rho(x)=F_{12}(x) \ ,
\end{equation}
it is similar to a magnetic field. 
On the lattice this magnetic field density 
is extracted from the plaquette, 
\begin{equation}
P(\vec n)=U(\vec n)_1U(\vec n+\hat 1)_2U(\vec n+\hat 2)_1^*U(\vec n)_2^*
\end{equation}
and a possible definition of topological charge
density is,  
\begin{equation}
\rho_1(\vec n)={1 \over 2 \pi} \arg P(\vec n) \ ,
\label{topo density}
\end{equation}  
where $-1/2 < \rho_1(\vec n) < 1/2$.
This definition of the topological density is not 
unique, the necessary condition is that it must 
reproduce the correct continuum limit. 
The definition 
(\ref{topo density}) of the 
topological charge is particularly interesting because
it is quantized.
It is clear that the total topological charge
of the lattice, which is defined with, 
\begin{equation}
q= \sum_{\vec n} \rho(\vec n) \ ,
\end{equation}
is an integer.
The charge $q_1$ is the sum of the decimal
part of a set of numbers, and these numbers
have a vanishing sum, therefore $q_1$ is an integer.
This is similar to $d=2$ compact QED, where the 
magnetic monopoles arise naturally on the lattice.
Here the topological charge is equivalent to a 
magnetic flux through the toroidal $d=2$ lattice.
\par
We now construct non-trivial gauge configurations.
In this work we are particularly interested
in extrapolating continuum properties to
finite lattices. 
We specialize to gauge configurations
with uniform topological charge density.
These configurations are unique, except for
gauge transformations.
In the $N \times N$ lattice a simple configuration 
is built with the prescription \cite{Chiu-testexact} 
of Chiu,
\begin{eqnarray}
U_j(\vec n) &=&\exp \left[ i A_j(\vec n) \right] \ ,
\nonumber \\
A_1(n_1,n_2)&=&-2 \pi \, Q{ n_2 -1 \over N^2} \ , 
\nonumber \\ 
A_2(n_1,N)&=&2 \pi \, Q{ n_1 -1 \over N} \ ,
\label{gauge}
\end{eqnarray}
where the undefined $A_j(n)$ are zero. 
With this definition (\ref{gauge}) and in the particular 
case of integer $Q$, the topological density is constant.
If $-N^2/2 <Q <N^2 /2$ we find that
$\rho_1=Q/N^2$ and 
that the topological charge is $q_1=Q$.
\par
The case where Q is an arbitrary continuous parameter
can be used when we are interested in an
interpolation between the cases  of
uniform density. Then the density is uniform,
except at the single point (N,N). The
gauge configuration (\ref{gauge}) is
periodic in $Q$, with a period of $N^2$.
In the relevant range $-N^2/2 <Q <N^2 /2$ 
we find that the topological
charge $q_1$ of eq. (\ref{topo density}) is a 
step-like function of $Q$,
\begin{equation}
q_1=\mathop{\mathrm{int}}
\left(Q {N^2-1 \over N^2}+{ 1 \over 2} \right) \ .
\label{step}
\end{equation}
\par

\section{The Wilson action}
\label{sec-Wilson}
\par
The Wilson \cite{Wilson} action 
is the simplest and most widely
used action in lattice field theory.
It also provides a good example
to observe roots and doubles,
\begin{eqnarray}
^w\hspace{-.05cm}D &=& {1 \over a} \sum_j \gamma_j C_j + 
{2 r \over a} \sum_j {B_j} \ ,
\nonumber \\
 {C_j}_{\vec n ,\vec n'}  
&=&{
U(\vec n)_j\delta_{n+\hat j,n'}-U(\vec n')_j^\dagger\delta_{\vec n,\vec n'+\hat j} 
\over 2 } \ ,
\nonumber \\
{B_j}_{\vec n ,\vec n'} 
&=& {2 \delta_{\vec n,\vec n'}- 
U(\vec n)_j\delta_{n+\hat j,\vec n'}-U(\vec n')_j^\dagger
\delta_{\vec n, \vec n'+\hat j} \over 4} \ ,
\label{wilsoneg}
\end{eqnarray}
where $r$ is a parameter of the order of $1$. 
The vector $\gamma_j$ term of eq. (\ref{wilsoneg}) is the naive massless 
Dirac action on the lattice, where the derivative $C_j$ is computed by means of
finite differences. 
In the limit of free fermions, which corresponds to $U_j=1$,
all these operators commute. In this 
case the Fourier transform of the above matrices depend on a 
single momentum,
\begin{eqnarray}
\label{trigonometric}
I(k) &=& 1 \ ,
\nonumber \\
C_j(k) &=&  i\sin k_j \ ,
\nonumber \\
B_j(k) &=& \sin^2 {k_j \over 2} \ .
\end{eqnarray}
In the free case, and in  $d$ dimensions, $C_j(k)$ 
has $2^d$ roots at all possible
combinations of $k_j=0$ and $k_j=\pi$ , see eq. (\ref{trigonometric}). 
This is the source of the doubler problem because only the root at 
vanishing $k$ is physical. 
Wilson \cite{Wilson} included the scalar $B_j$ term in eq. (\ref{wilsoneg})
to remove the unwanted doubles.
\par
However this scalar term $B_j$ is not chiral invariant. Chiral
invariance can be implemented with the L\"uscher transformation.
However Horvath showed that the $RD$ operator can at most \cite{Horvath}
be exponentially local, therefore the $^w\hspace{-.05cm}R$ of the GWR
is not a simple number in this case. From the GWR relation, 
see eq. (\ref{GWR 5}) we get, 
\begin{equation}
^{w}\hspace{-.05cm}D
^{w}\hspace{-.05cm}R=I+ \, ^{w}\hspace{-.05cm}D {1\over ^{w}\hspace{-.05cm}D^\dagger} ,
\label{simple again}
\end{equation} 
where the Wilson action $^{w}\hspace{-.05cm}D^\dagger$ is 
in general invertible, see Section  \ref{sec-eigen Wilson}.
We can remark that the operator in eq. (\ref{simple again})  
is a solution of the GWR, however it is
not exponentially local. 
In this case the nontrivial operator of the L\"uscher 
transformation,
\begin{equation}
\gamma_5\left( I-{1\over 2}
\, ^{w}\hspace{-.05cm}R
^{w}\hspace{-.05cm}D \right)= {1\over 2}\left( \gamma_5 
+ ^{w}\hspace{-.05cm}D^ {-1}\gamma_5 ^{w}\hspace{-.05cm}D \right),
\end{equation}
is also non-local. Moreover 
$^{w}\hspace{-.05cm}D ^{w}\hspace{-.05cm}R$ is not $\gamma_5$ Hermitean.
This excludes 
both the Wilson Action and the $^{w}\hspace{-.05cm}D ^{w}\hspace{-.05cm}R$
as the best candidate to study exact chiral symmetry on the lattice.

\section{Eigenvalues of the Wilson action}
\label{sec-eigen Wilson}
Here we depart from the free case and study the 
eigenvalues of the Wilson action for topological gauge
configurations . This continues reference \cite{Farchioni}.
The Wilson action  $^w\hspace{-.05cm}D$ is the simplest 
and most used lattice action, see Section \ref{sec-Wilson}.
Our framework is the massless Wilson action
with parameters $a=r=1$, see eqs 
(\ref{wilsoneg}), in a simple 2-dimensional
U(1) gauge theory which is a discrete version
of the Schwinger model, see Section \ref{sec-topology}.  
In the continuum the Wilson action \cite{Hernandez}
exactly complies with the Index Theorem (\ref{index}), 
but for finite lattices it does not.
Our aim in this section is to extrapolate the topological charge  
from the continuum to the lattices of different size $N \times N$,
and to study how the Wilson Action deviates from the Index theorem. 
\par
We encounter the problem that the topological charge and
the topological density on the lattice are not uniquely defined. 
This problem is not present in the continuum, 
where the topological density and its derivatives are well defined.
This implies that a lattice action close to the continuum limit is only
allowed to have a topological charge density both small and nearly 
constant.  However we are interested, both 
physically and mathematically, in extrapolating the Index Theorem to the
case of finite lattices, with finite topological charge.
And then we want to investigate how far the topological 
density can be increased without spoiling the index theorem. 
To remain as close as possible to the continuum definitions, 
we momentarily specialize to very smooth gauge configurations. 
In particular we study the Wilson action in configurations with 
constant plaquette density, see Section \ref{sec-topology}, 
which is equivalent to a constant topological charge density.
\par
\begin{figure}
\begin{picture}(280,540)(0,0)
\put(-60,360){\begin{picture}(0,0)(0,0)
\put(10,10){\includegraphics{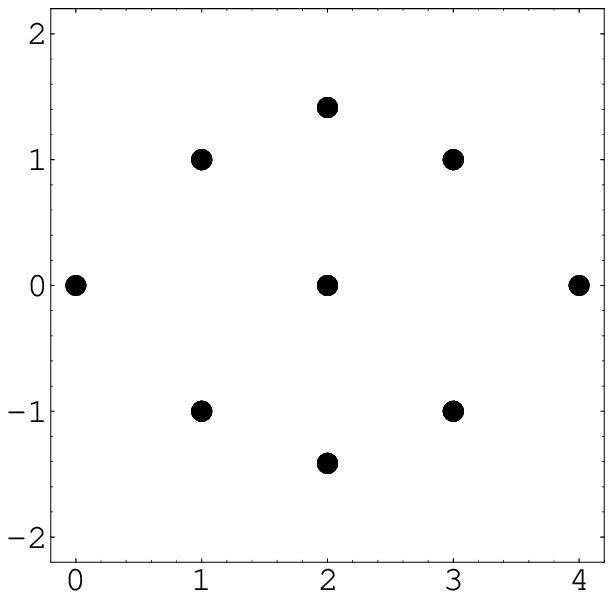}}
\put(30,170){Im$(\lambda)$}
\put(150,30){Re$(\lambda)$}
\put(140,170){$_{Q=0.0}$}
\end{picture}}
\put(140,360){\begin{picture}(0,0)(0,0)
\put(10,10){\includegraphics{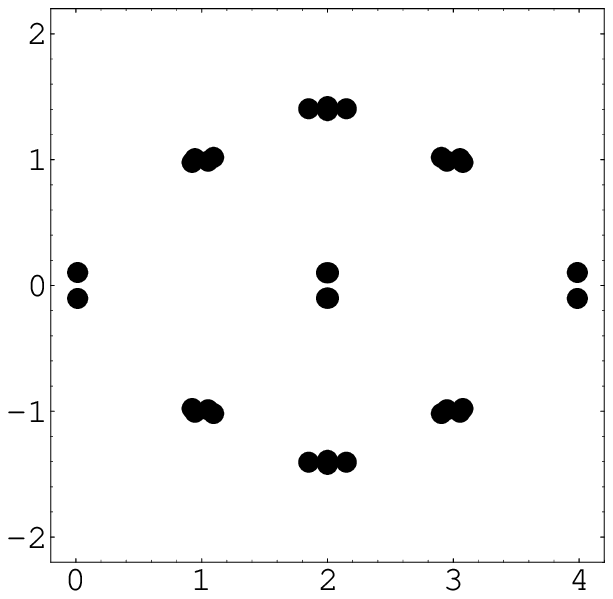}}
\put(30,170){Im$(\lambda)$}
\put(150,30){Re$(\lambda)$}
\put(140,170){$_{Q=0.125}$}
\end{picture}}
\put(-60,180){\begin{picture}(0,0)(0,0)
\put(10,10){\includegraphics{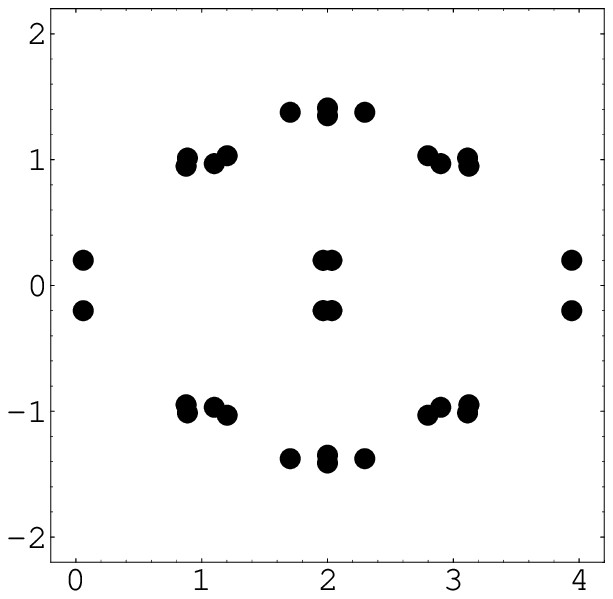}}
\put(30,170){Im$(\lambda)$}
\put(150,30){Re$(\lambda)$}
\put(140,170){$_{Q=0.25}$}
\end{picture}}
\put(140,180){\begin{picture}(0,0)(0,0)
\put(10,10){\includegraphics{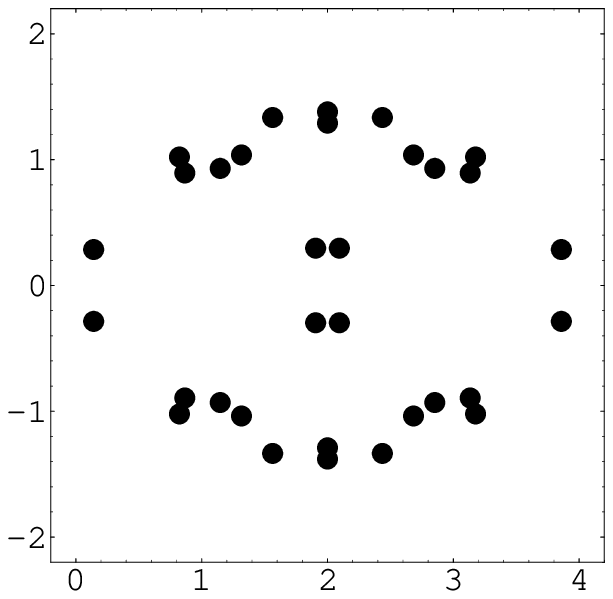}}
\put(30,170){Im$(\lambda)$}
\put(150,30){Re$(\lambda)$}
\put(140,170){$_{Q=0.375}$}
\end{picture}}
\put(-60,00){\begin{picture}(0,0)(0,0)
\put(10,10){\includegraphics{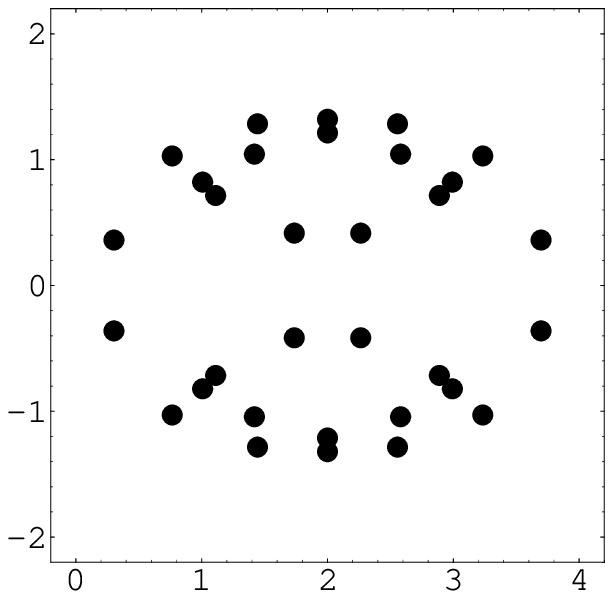}}
\put(30,170){Im$(\lambda)$}
\put(150,30){Re$(\lambda)$}
\put(140,170){$_{Q=0.5}$}
\end{picture}}
\put(140,0){\begin{picture}(0,0)(0,0)
\put(10,10){\includegraphics{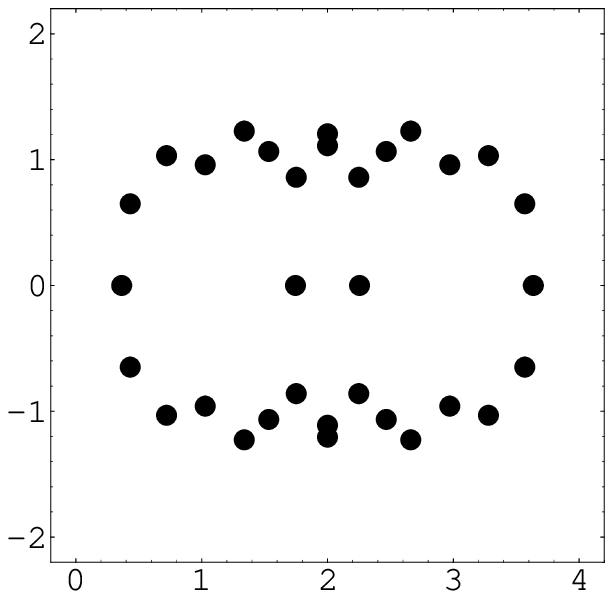}}
\put(30,170){Im$(\lambda)$}
\put(150,30){Re$(\lambda)$}
\put(140,170){$_{Q=0.625}$}
\end{picture}}
\end{picture}
\caption{Sequence of the eigenvalues of the Wilson action
as a function of the topological parameter $Q$.
We show the Argand plot of the eigenvalues $\lambda$ of the 
Wilson $^wD$, on a $4 \times 4$ lattice and with parameters $a=r=1$.
}
\label{eigenwil1}
\end{figure}
\begin{figure}
\begin{picture}(280,540)(0,0)
\put(-60,360){\begin{picture}(0,0)(0,0)
\put(10,10){\includegraphics{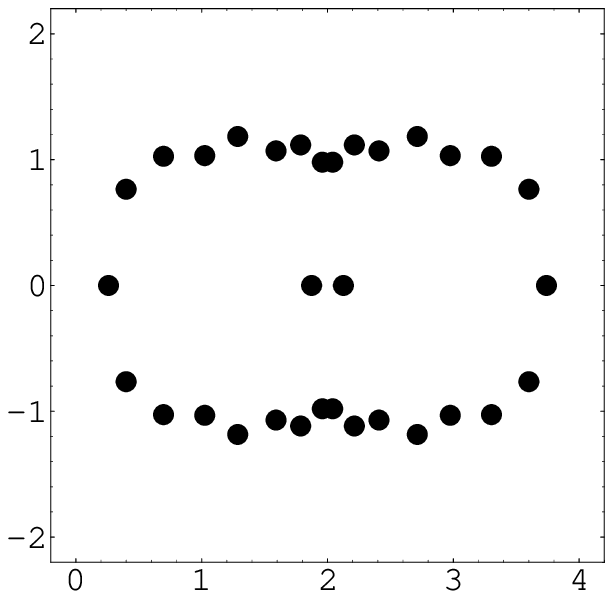}}
\put(30,170){Im$(\lambda)$}
\put(150,30){Re$(\lambda)$}
\put(140,170){$_{Q=0.75}$}
\end{picture}}
\put(140,360){\begin{picture}(0,0)(0,0)
\put(10,10){\includegraphics{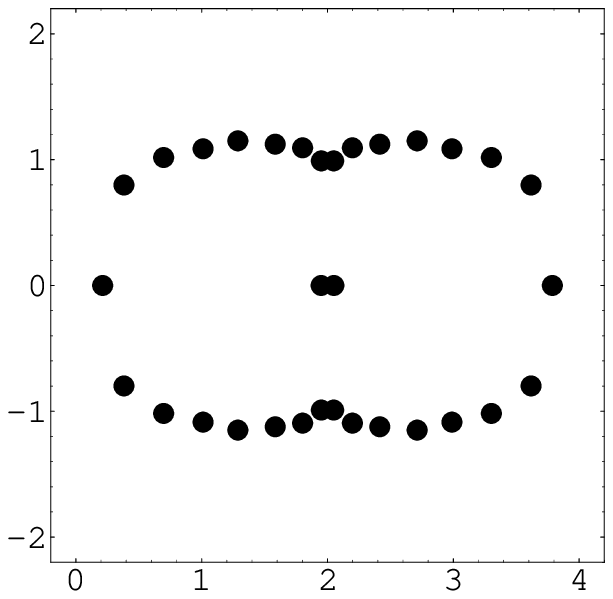}}
\put(30,170){Im$(\lambda)$}
\put(150,30){Re$(\lambda)$}
\put(140,170){$_{Q=0.875}$}
\end{picture}}
\put(-60,180){\begin{picture}(0,0)(0,0)
\put(10,10){\includegraphics{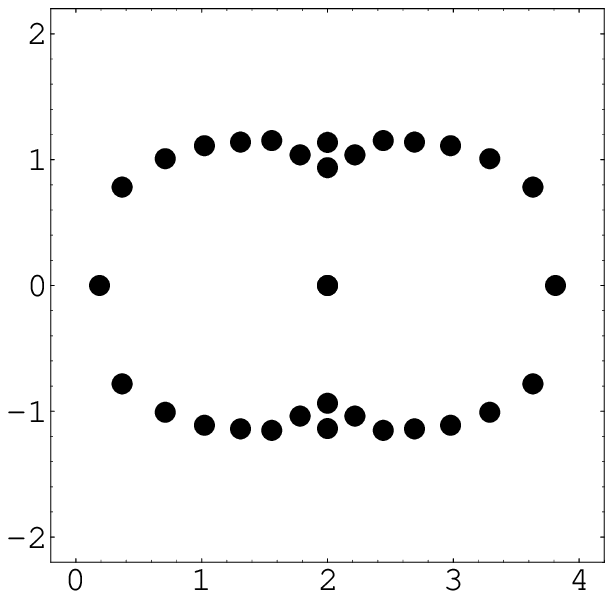}}
\put(30,170){Im$(\lambda)$}
\put(150,30){Re$(\lambda)$}
\put(140,170){$_{Q=1.0}$}
\end{picture}}
\put(140,180){\begin{picture}(0,0)(0,0)
\put(10,10){\includegraphics{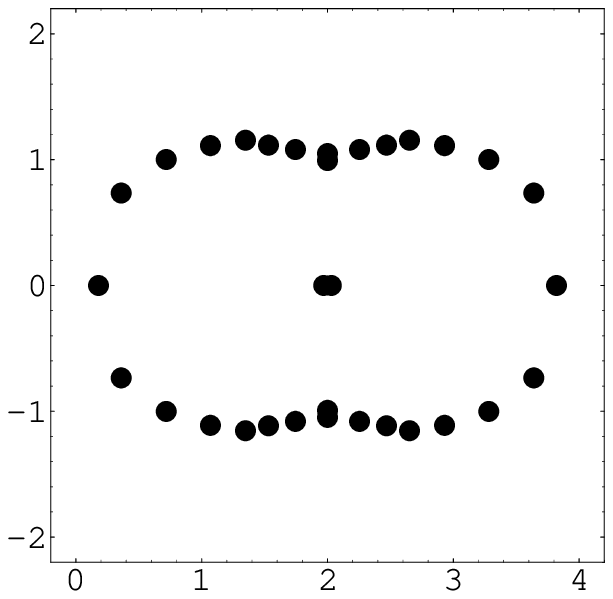}}
\put(30,170){Im$(\lambda)$}
\put(150,30){Re$(\lambda)$}
\put(140,170){$_{Q=1.125}$}
\end{picture}}
\put(-60,00){\begin{picture}(0,0)(0,0)
\put(10,10){\includegraphics{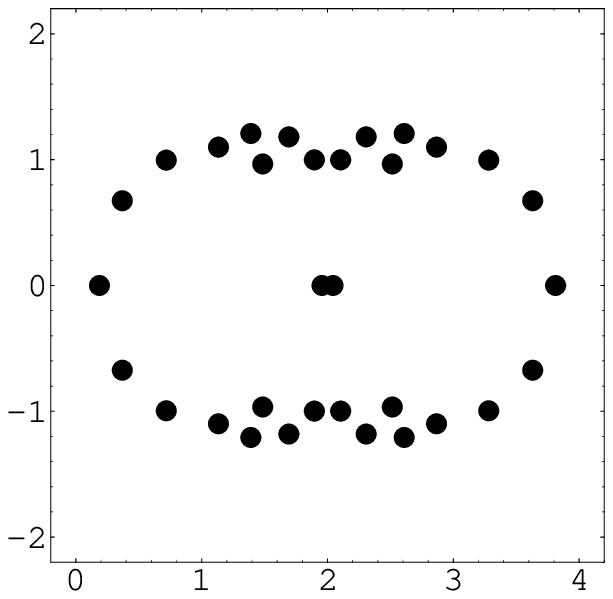}}
\put(30,170){Im$(\lambda)$}
\put(150,30){Re$(\lambda)$}
\put(140,170){$_{Q=1.25}$}
\end{picture}}
\put(140,0){\begin{picture}(0,0)(0,0)
\put(10,10){\includegraphics{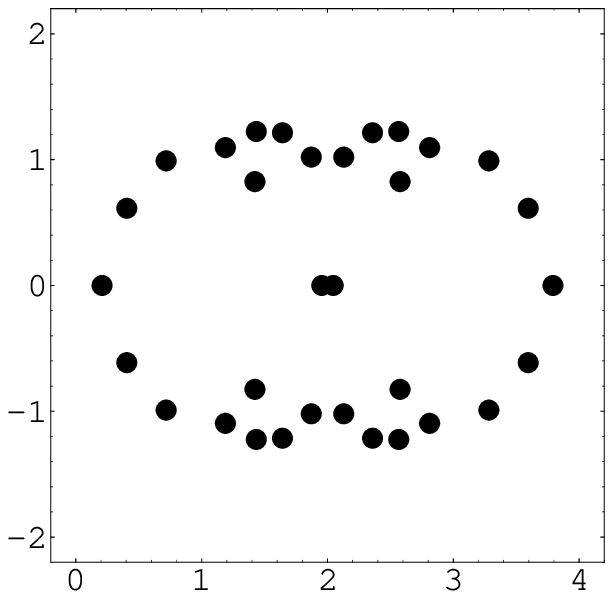}}
\put(30,170){Im$(\lambda)$}
\put(150,30){Re$(\lambda)$}
\put(140,170){$_{Q=1.375}$}
\end{picture}}
\end{picture}
\caption{Sequence of eigenvalues of the Wilson action, 
continuing Fig. 2.
The Wilson action successfully separates the doubling of the spectrum, 
not only in the free case of $Q=0.0$ , but also in topologically nontrivial cases. 
}
\label{eigenwil2}
\end{figure}
\begin{figure}
\begin{picture}(280,540)(0,0)
\put(-60,360){\begin{picture}(0,0)(0,0)
\put(10,10){\includegraphics{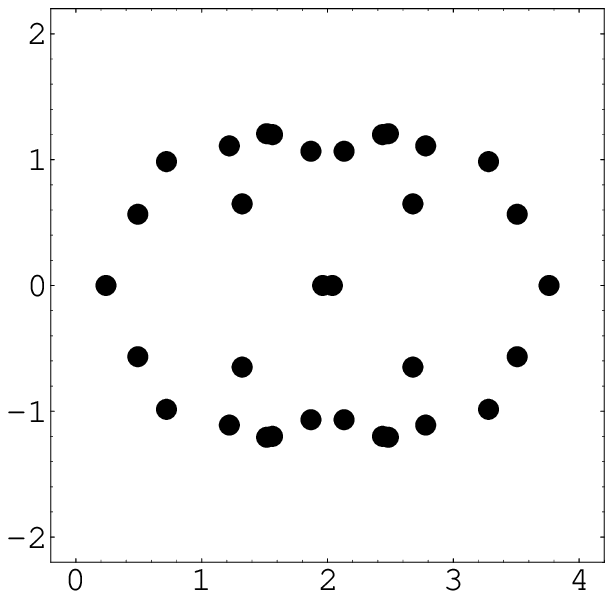}}
\put(30,170){Im$(\lambda)$}
\put(150,30){Re$(\lambda)$}
\put(140,170){$_{Q=1.5}$}
\end{picture}}
\put(140,360){\begin{picture}(0,0)(0,0)
\put(10,10){\includegraphics{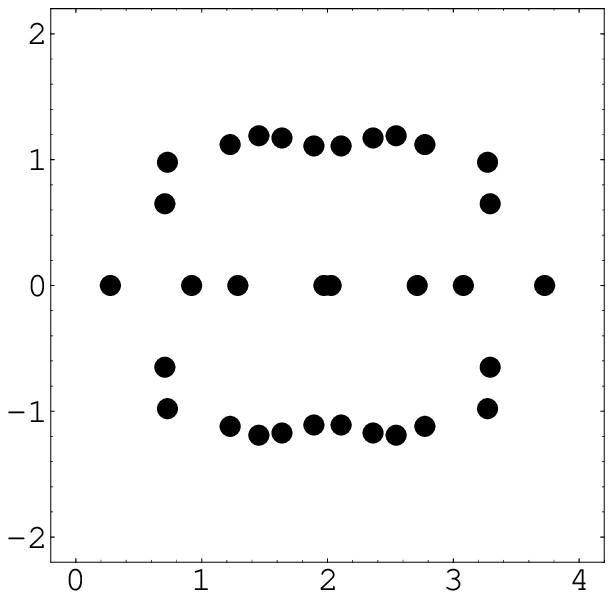}}
\put(30,170){Im$(\lambda)$}
\put(150,30){Re$(\lambda)$}
\put(140,170){$_{Q=1.625}$}
\end{picture}}
\put(-60,180){\begin{picture}(0,0)(0,0)
\put(10,10){\includegraphics{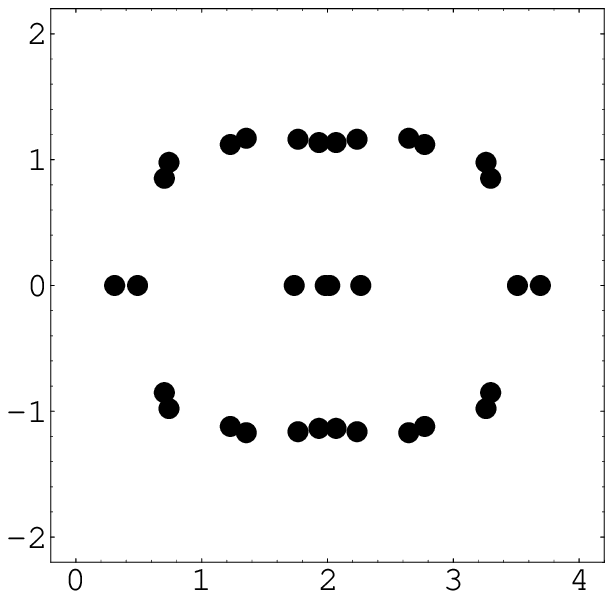}}
\put(30,170){Im$(\lambda)$}
\put(150,30){Re$(\lambda)$}
\put(140,170){$_{Q=1.75}$}
\end{picture}}
\put(140,180){\begin{picture}(0,0)(0,0)
\put(10,10){\includegraphics{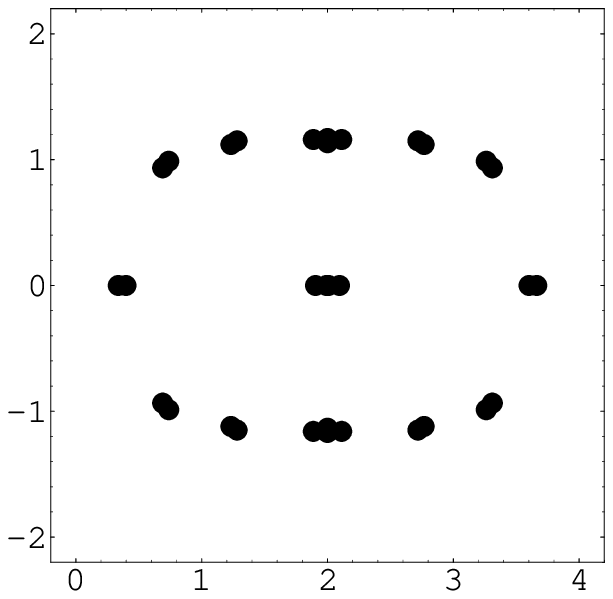}}
\put(30,170){Im$(\lambda)$}
\put(150,30){Re$(\lambda)$}
\put(140,170){$_{Q=1.875}$}
\end{picture}}
\put(-60,00){\begin{picture}(0,0)(0,0)
\put(10,10){\includegraphics{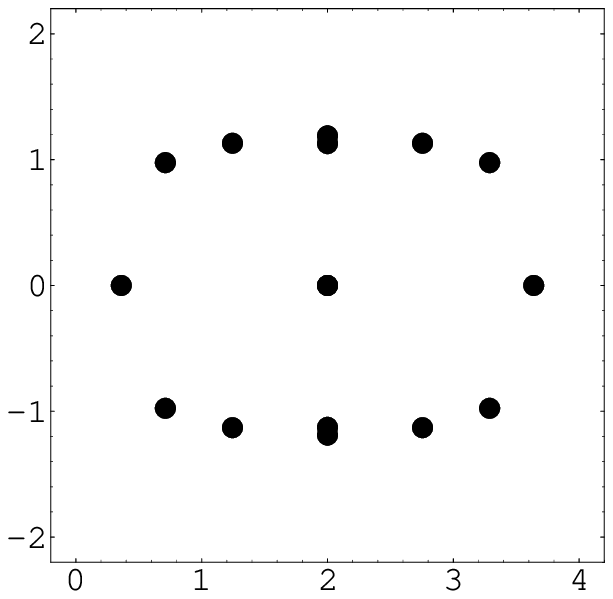}}
\put(30,170){Im$(\lambda)$}
\put(150,30){Re$(\lambda)$}
\put(140,170){$_{Q=2.0}$}
\end{picture}}
\put(140,0){\begin{picture}(0,0)(0,0)
\put(10,10){\includegraphics{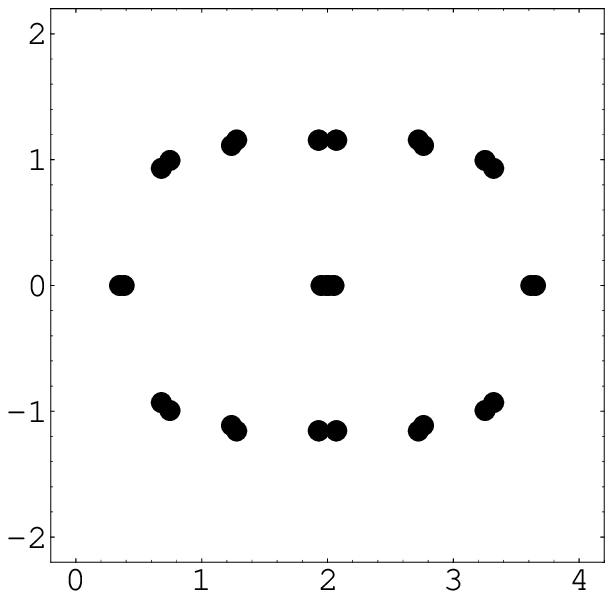}}
\put(30,170){Im$(\lambda)$}
\put(150,30){Re$(\lambda)$}
\put(140,170){$_{Q=2.125}$}
\end{picture}}
\end{picture}
\caption{Sequence of eigenvalues of the Wilson action, 
continuing Figs. 1 and 2.
For some particular values of $Q$, two opposite pairs of complex eigenvalues 
transform into two opposite pairs of real eigenvalues.
}
\label{eigenwil3}
\end{figure}

First we study some general properties of the Wilson action.
Because it is $\gamma_5$ Hermitean, the spectrum of its eigenvalues, 
here denoted $\lambda$, is symmetric with respect to the real axis of 
the Argand plot, see Section \ref{sec-properties}.
Moreover in $d=2$ it turns out that in the case of even lattices, 
the spectrum of $^w\hspace{-.05cm}D-2I$ in the Argand plot is 
also symmetric with respect to the imaginary axis.  The transformation 
$(\lambda-2) \rightarrow - (\lambda-2)^*$ leaves the spectrum 
invariant (see for instance Figs. \ref{eigenwil1},\ref{eigenwil2},\ref{eigenwil3}).
So we will use even $N \times N$ lattices with $N\ge 4$ from now on, 
for simplicity.
We also remark that $0 < $ real$(\lambda) < 4$ for any non vanishing
gauge configuration, thus $^w\hspace{-.05cm}D$ is invertible
(except in the free quark case).
In what concerns $\gamma_5 (^w\hspace{-.05cm}D-2I)$ we observe
that its eigenvalues are exactly symmetric with respect to
the origin. This implies for instance that the traces  
tr$\left[ \gamma_5 \, (^w\hspace{-.05cm}D-2I)^{-1} \right]$
and 
tr$\left[ \gamma_5 \, ^w\hspace{-.05cm}D \right]$
both vanish for any finite $N$. This is not the
case of the trace tr$\left[ \gamma_5 \, 
^w\hspace{-.05cm}D^{-1} \right]/{\cal Z}$ 
which is non vanishing and has been proposed as a topological index
\cite{Bali}, where ${\cal Z}$ is a constant normalizing factor.
However we will not further study this proposed topological charge
because it clearly is neither local nor integer.
\par
We now study quantitatively the real eigenvalues and the 
topological properties of the Wilson action
$^w\hspace{-.05cm}D$. 
We are interested in 
eigenvalues $\lambda \simeq 0$ with chirality $\chi \simeq \pm 1$, 
see Section \ref{sec-properties}.
The non vanishing chirality implies that the relevant
eigenvalues are the real eigenvalues. 
The $B_j$ term was introduced in the free action by Wilson 
to separate the four roots of the $C_j$, 
see Section \ref{sec-Wilson}. In the free limit this 
results in four real eigenvalues: one root at $0$ , 
two degenerate eigenvalues equal to $2$ 
and one eigenvalue equal to $4$.  
Here we find that quite a similar separation exists
in the quadruplets of real eigenvalues that occur in 
gauge configurations with a finite topological charge
 (see Figs. \ref{eigenwil1},\ref{eigenwil2},\ref{eigenwil3}).
Moreover the chirality of these real eigenvalues 
approximately agree with the Index Theorem.
In particular we observe that,
\\
- The number of real eigenvalues is always a multiple of four,
$4|q_2|$. The integer $q_2=-$ sign$(\chi)|q_2|$ is a possible 
definition of the topological charge of the gauge configuration, 
where $\chi$ is the chirality of the smallest real eigenvalue.
\\
- In the present case of a constant topological 
density, $q_2=q_1$ if $|\rho_1|$ is smaller than half of 
the maximum possible charge density. The integer topological
charge $q_1$ and its topological density 
$\rho_1$ are defined in Section \ref{sec-topology}.
\\
- The real eigenvalues and the corresponding 
chiralities are close to integers. The difference
is proportional to a small number, $\epsilon = | q_2 |/ N^2 $.
The $4|q_2|$ eigenvalues can be divided in the following 3 sets of 
 eigenvectors, with eigenvalues,
\begin{eqnarray} 
\lambda= 0 + 3.0 \epsilon + o(\epsilon^2) \ &,& \ 
\chi =  [ -1 + o(\epsilon^2) ] sign(q_2) ,
\nonumber \\
\lambda= 2 + o(\epsilon^2) \ &,& \ 
\chi =  [ 1 -2.0 \epsilon + o(\epsilon^2) ] sign(q_2) ,
\nonumber \\
\lambda= 4 - 3.0 \epsilon + o(\epsilon^2) \ &,& \ 
\chi =  [ -1 + o(\epsilon^2) ] sign(q_2) ,
\end{eqnarray}
which contain respectively $|q_2|$, $2|q_2|$ and $|q_2|$ eigenvalues.
\\
- When $|q_2|$ approaches the maximum value of the order of $n^2/4$, 
these 3 sets of eigenvalues spread out to 3 finite intervals, 
and eventually they overlap. 
It turns out that right before these intervals overlap,
they just leave a gap at $[1.0,1.2]$ .
\\
- In the continuum limit of $N \rightarrow \infty$, and 
for very smooth topological objects, all the 
different $\epsilon \rightarrow 0$. 
So we find that in this limit of large N, 
and from the view point of zero
modes, the Index Theorem  is verified,
\begin{equation}
n_- - n_+ =q_2 \ .
\label{yesindex}
\end{equation}
This result agrees with the proofs of 
Fujikawa \cite{Fujikawa} and Adams \cite{Adams}.
\par
If we now leave the case of constant topological density, 
it occurs that $|\rho_1|< 0.22 \,$ 
(we mean that the absolute value of the topological density 
$\rho_1$ is smaller than 0.22 in any plaquette of the lattice)
is a sufficient condition to enforce that $q_2=q_1$.
We verified this crucial result with a large number of randomly 
generated gauge configurations, for all possible $q_1$.
All the remaining properties, that we just detailed, of the eigenvalues 
of the Wilson action are also maintained when $|\rho_1|< 0.22$ . 
This suggests that it is possible to extrapolate the continuum
limit to a significant and finite subinterval of the range 
$-1/2<\rho_1< 1/2$.
\par
To illustrate how the complex eigenvalues
transform in real ones, we arbitrarily
interpolate between different constant 
charge densities with a continuous 
variation of the topological parameter $Q$ 
( see Section \ref{sec-topology} ).
It turns out that for some particular values of $Q$,
which produce a large  $|\rho_1|$ at the point $(N,N)$,
two opposite pairs of
complex eigenvalues (with $0$ chirality)
quite suddenly transform into two opposite pairs of 
real eigenvalues, and then these real eigenvalues
continuously increase their chirality $|\chi|$ .
This transition point is continuous, but at
this precise value of $Q$ the velocity of the
eigenvalues in the Argand plot is infinite.
At this point the topological number $q_2$
steps up (down). The opposite also happens, 
at some other particular values of $Q$, where $4$ 
real eigenvalues suddenly transform into complex
eigenvalues.
The different topological charges can also be computed
( see Section \ref{sec-topology} ).
and the observed pattern is quite general. The  
charge $q_2$, defined with the real eigenvalues of the
Wilson action, 
and the charge $q_1$, defined with
the plaquette density $P$ in Section \ref{sec-topology},
are quite close only up to half of the maximum possible 
topological $q_1$. This verifies our empirical rule of $|\rho_1|<0.22$.
\par
In Figs. \ref{eigenwil1},\ref{eigenwil2},\ref{eigenwil3} we show 
a sequence of the eigenvalues of the Wilson action for consecutive
values of the interpolating topological parameter $Q$.
A simple $4 \times 4$ lattice is used and the eigenvalues are 
displayed in the Argand plan.
 
\section{Criterion}
\label{sec-criterion}
\par
We aim to understand how the Index Theorem 
extrapolates from the continuum limit to finite 
lattices with finite topological charge densities.
In d=2, and providing that the topological 
density $|\rho_1|<0.22$, we show that the Wilson action
$^{w}\hspace{-.05cm}D$ 
has the correct number of small real eigenvalues 
to comply with the index theorem. The problem is that
these eigenvalues are not exactly vanishing and that
their chirality is not exactly $\pm 1$. The deviations
to the index theorem are of order $\epsilon=|q_1| / N^2$,
where $q_1$ is the integer topological charge.
We also prove that the GWR Neuberger solutions either project 
these approximate zero modes of $^{w}\hspace{-.05cm}D$
on the origin, with correct chirality, or send them close
to the remote other end $2/R$ of the unitary circle.
\par
This motivates a new criterion for GWR solutions  $^{gwr}\hspace{-.05cm}D$
with index identical to $q_1$  ( and $\gamma_5$ Hermitean, constructed 
with the Wilson action ), which relies in the function $f(\lambda)$.
This function $f(\lambda)$ is straightforwardly defined by replacing
$^{w}\hspace{-.05cm}D$ and $^{w}\hspace{-.05cm}D^\dagger$ by
the same real  number $\lambda$ in the expression for 
$^{gwr}\hspace{-.05cm}D$. 
We map $^{gwr}\hspace{-.05cm}D(^{w}\hspace{-.05cm}D,^{w}\hspace{-.05cm}D^\dagger)
\rightarrow f(\lambda)=^{gwr}\hspace{-.05cm}D(\lambda,\lambda)$.
\par
The $f(\lambda)$ is a trivial case of a GWR solution. Moreover the $\gamma_5$
hermiticity implies that $f(\lambda)$ is real. We thus conclude
that $f(\lambda)=0$ or $f(\lambda)=2/R$. 
\par
In the free case $^{w}\hspace{-.05cm}D$ and 
$^{w}\hspace{-.05cm}D^\dagger$ commute, and have the correct
real eigenvalues at $0$, while the doubles correspond to 
the eigenvalues $2$ or $4$. We remark that the doubles correspond to 
pathological eigenvectors, with alternating signs in neighboring
lattice sites. We conclude that, unless the fermion fields are 
redefined, a correct GWR solution must produce $f(0)=0$,   
$f(2)=2/R$ and  $f(4)=2/R$.
\par
Studying small topological densities  $\rho_1$ with finite $q_1$, we can further 
show that $f([0,  3.0 \epsilon])=0$, $f(2)=2/R$,  $f([4-3.0 \epsilon,4])=2/R$. 
In the case of small topological densities, $^{w}\hspace{-.05cm}D$ and  
$^{w}\hspace{-.05cm}D^\dagger$ have the same eigenvalues, but they do 
not exactly have the same eigenvectors. 
Nevertheless the relevant small real eigenvalues $\lambda$ 
of the Wilson action, have a chirality 
$ \chi =  [ -1 + o(\epsilon^2) ]$sign$(q_2) $
and this implies that the difference between the 
$^{w}\hspace{-.05cm}D$ eigenvector
$v$ and the $^{w}\hspace{-.05cm}D^\dagger$ eigenvector
$\gamma_5 v$ is very small,
of order $\epsilon^2$. 
This shows that $
^{gwr}\hspace{-.05cm}D \, v = f(\lambda) v + o(\epsilon^2)
$, the correct eigenvector $v'$ of
$^{gwr}\hspace{-.05cm}D$ is quite close to $v$ (and to $\gamma_5 v$ ). 
{\em 
If $^{gwr}\hspace{-.05cm}D$ complies with the index theorem,
}
we can assume that this eigenvector $v'$ is a zero mode, because
$v$ is already quite close to the correct fermionic zero mode.
Therefore $f(\lambda)\simeq 0$. 
Moreover we can only have $f(\lambda)= 0 $ or 
$2/R$. This implies that $f(\lambda)=0$. 
Because $\lambda= 0 + 3.0 \epsilon + o(\epsilon^2)$ we find that 
{\em 
$f$ projects the interval $[0,  3.0 \epsilon]$
on the origin $0$ . 
Inversely, if $f(\lambda)=0$ ,
}
$^{gwr}\hspace{-.05cm}D$ has $|q_1|$ eigenvectors close to $f(\lambda)=0$. 
Moreover they all have similar chirality 
$ \chi =  [ -1 + o(\epsilon^2) ]$sign$(q_2) $. Even if these eigenvectors mix, which 
is natural because they are very close, their chirality will remain close to 
sign$(q_2)$, and finite. 
{\em 
Therefore these $|q_1|$ eigenvalues are real, 
vanishing, with the correct chirality, and in the right
number to comply with the index theorem 
}. 
Similarly we can show that $f$ projects
on $2/R$ both a small neighborhood of the point $2$ 
and the interval $[4-3.0 \epsilon,4]$. 
If we assume that the complex eigenvalues of the Wilson action
are irrelevant to the index because they have vanishing chirality,
we conclude that the  
condition $f([0,  3.0 \epsilon])=0$, $f(2)=2/R$,  $f([4-3.0 \epsilon,4])=2/R$ 
is both a necessary and a sufficient condition for any 
$\gamma_5$ Hermitean GWR solution, constructed with the Wilson action, 
and free from doubling, to comply with the index theorem in the case
of small topological density. 
\par
In the case of a finite topological charge density the 
three sectors of real eigenvectors of the Wilson that include  
$0$, $2$ and $4$ spread out. Analytically it is hard to
find how far the exact criterion that we just derived for 
small $\epsilon$ can be extended to finite $\epsilon$.
Nevertheless our numerical study of the Wilson action 
shows that up to moderate densities $|\rho_1|<0.22$, 
the sectors including $0$, $2$ and $4$ remain separated. 
For instance the sector of $0$, with chiralities
$\chi \simeq -$ sign$(q_1)$, and the sector of $2$, 
with chiralities $\chi \simeq +$ sign$(q_1)$, are separated
by a gap located at $\lambda_0=1.1 \pm 0.1$ . It is thus quite 
plausible that a correct GWR solution should project
the smaller $|q_2|$ real eigenvalues $\lambda$
of $^{w}\hspace{-.05cm}D$ on
$0$, and that the remaining $3|q_2|$ real eigenvalues
should be projected onto the other end of the circle, 
close to $2/R$. 
We may now determine which
is the function $f(\lambda)$ that maximally complies with the
index theorem, up to the highest possible topological
density $0.22$ . This $f(\lambda)$ is a step function that jumps 
precisely at the gap $\lambda_0=1.1 \pm 0.1$.
\par
Finally when the density is large, $|\rho_1|>0.22$ or more
at one or many plaquettes of the lattice, the index theorem is 
spoiled. This is natural on the lattice, because for large 
densities the different definitions of topological charge 
density diverge and the topological charge is not conserved.
In what concerns the three sectors of real eigenvectors of the 
Wilson action, that include respectively $0$, $2$ and $4$, 
they swell so much that they start to overlap. 
The real eigenvalues either mix and transform into 
complex eigenvalues (with vanishing chirality), or they 
carry their chirality to the wrong sector. 
We remark that this is not a physical problem, 
providing the gauge configurations with large plaquette 
densities are suppressed by the gauge part of the action.
\par
We now present our criterion, which is very simple to use. 
{\bf ``
A $\gamma_5$ Hermitean GWR solution $^{gwr}\hspace{-.05cm}D$ with 
constant $R$ and constructed with the Wilson $^{w}\hspace{-.05cm}D$
action complies maximally with the index theorem if and only if 
$^{gwr}\hspace{-.05cm}D \rightarrow {2 \over R} \theta(\lambda-\lambda_0)$
when we replace  $^{w}\hspace{-.05cm}D \, , \,
^{w}\hspace{-.05cm}D^\dagger \rightarrow \lambda$ ,
where $\lambda$ is a real number, $0 \leq \lambda \leq 4$ and 
$\lambda_0$ belongs to a well determined and narrow subinterval 
of $]0,2[$
''}.
Our criterion also applies to GWR solutions constructed with any
other action $^{o}\hspace{-.05cm}D$ (other than the Wilson action), 
that approximately complies with the index 
theorem and that is $\gamma_5$ hermitian,
 $^{o}\hspace{-.05cm}D^\dagger=
\gamma_5 \, ^{o}\hspace{-.05cm}D \, \gamma_5$.

\section{Conclusion}
\label{sec-conclusion}
\par
This talk is devoted to clarify how far 
the Index Theorem can be extrapolated from the continuum 
to finite lattices with finite topological charge
densities.
\par
The $d=2$ Wilson action $^{w}\hspace{-.05cm}D$ is examined, 
and we find that it approximately complies with the Index 
Theorem for finite topological charge densities lower  
$|\rho_1|=0.22$.
We also study the GWR solution of Neuberger which
exactly complies with the Index Theorem.
Finally we produce a criterion for GWR solutions,
constructed with the Wilson action, that maximally 
comply with the Index Theorem.

\par
With our criterion, it is simple to show that
locality constitutes a sufficient condition for 
a Dirac action to comply with the index theorem
together with the simplest topological charge $q_1$.
It is clear that any local action is differentiable 
in the free limit, and therefore the condition
$f(0)=0\, , \ f(2)=2/R \, , \ f(4)=2/R$ can be extended
to a neighborhood of these points. Using our criterion,
this proves that any local action (GWR solution and 
with the correct free limit) complies with the index 
theorem at least  in the case of small topological charge 
densities. The locality conjecture is correct in the case of
actions built from the Wilson action, and it is 
probably correct in the general case. 
Moreover the locality condition can be improved. 
We aim at finding the action with the highest 
convergence radius in the neighborhood of the points 
$0, \, 2, \, 4$, and in a sense this is comparable
with finding the action with the highest locality.
 
\par
Our criterion complies with the Chiu and Zenkin criterion
\cite{Chiu-criterion} which states that the correct GWR solutions
must have at least one eigenvalue at $2/R$.
However we observe that it is possible to find a non-local 
action with eigenvalues at $2/R$, which does not have zero modes 
outside the free limit. 
Therefore, unlike the locality condition, the Chiu-Zenkin criterion 
is not a sufficient condition to force an action to comply with the 
Index Theorem, although it certainly constitutes a necessary condition. 

\par
Our criterion can be applied to the Neuberger GWR
solution defined with
$R\, ^n\hspace{-.05cm}D=I+ (^w\hspace{-.05cm}D-m_0) 
/ \sqrt{(^w\hspace{-.05cm}D-m_0)^\dagger
\, (^w\hspace{-.05cm}D-m_0)}$. 
With the $\lambda$ substitution we find that
$ ^n\hspace{-.05cm}D \rightarrow {2 \over R}
\theta(\lambda -m_0)$.
This complies with our criterion when $m_0=\lambda_0$,
and it is quite evident that this is the simplest
action which complies with it. The criterion shows that 
the Neuberger overlap action is the simplest GWR solution
constructed with the Wilson action that maximally complies 
with the Index Theorem. 

\par
Moreover, in the Schwinger model and using the 
dimensionless units of $a=r=1$, we find that 
$1.0 < \lambda_0 <1.2$. This determines 
$m_0=\lambda_0$ with a value which is not very far 
from the $m_0\simeq0.8$ that provides the highest
locality in the free limit.
This choice of $m_0$ increases the precision of the 
observation of Chandrasekharan \cite{Chandrasekharan-Schwinger}, 
stating that the Neuberger action only 
complies with the Index Theorem if $0< m_0 <2$,
($0< m_0 <2$ can also be derived from the root and pole 
structure of the free Neuberger action). 
The choice of $m_0$ completely
fixes the GWR parameter $R$ because the free
continuum limit of the lattice implies that $m_0= a/R$.
So the best choice seems to be $R \simeq 0.9 a$.

\par
Finally we outline possible continuations of this
study. 
We are researching the analytical extension of the proof of our  
criterion to finite topological densities $0<<|\rho_1|<0.22$. 
The repetition of the present study in four 
dimensions would also be physically relevant.

\par
I acknowledge Misha Polikarpov for explaining computing techniques
and topology on the lattice. I am also grateful to Herbert Neuberger 
for explaining the locality conjecture and for pointing to a
numerical error. I thank Dimitri Diakonov, Isabel Salavessa
and Emilio Ribeiro for discussions on the topological index.

%
%

\end{document}